# An FPGA Based energy correction method for one-to-one coupled PET detector: model and evaluation

Cong Ma[a,b,d], Xiaokun Zhao[a,b,c,d], Size Gao[a,b], Fengping Zhang[a,b], Guocheng Wu[a,b], Xing Li[a,b], Lei Lu[a,b], Hongwei Ye[a,b], Hua Qian[a,b]

a. *Minfound Medical System Co. Ltd. Yifeng Road, Hangzhou, 310018, China*
b. *Zhejiang MinFound Nuclear Medical Imaging Research Institute, Dongshan Road, Shaoxing, 312000, China*
c. *School of Medicine, Shaoxing University, Chengnan Road, Shaoxing, 312000, China*
d. *Modern Physics department, the University of Science and Technology of China. Jinzhai Road, Hefei, 230026, China*

*E-mail*: macong@mail.ustc.edu.cn, zhaoxk@mail.ustc.edu.cn

ABSTRACT: A PET scanner based on silicon photomultipliers (SiPMs) has been widely used as an advanced nuclear medicine imaging technique that yields quantitative images of regional in vivo biology and biochemistry. The compact size of the SiPM allows direct one-to-one coupling between the scintillation crystal and the photosensor, yielding better timing and energy resolutions than the light sharing methods that have to be used in photomultiplier tube (PMT) PET systems. To decrease the volume of readout electronics, a front-end multiplexer with position decoder is a common choice for the one-to-one system without a highly integrated application specific integrated circuit (ASIC). However, in this case we cannot measure each crystal's deposited energy inspired by an annihilation photon, so the inter-crystal scatter (ICS) events will lead to the crystal mispositioning and then deteriorate the detector intrinsic resolution. Besides, considering the events rejection within the energy window resulting from the gain dispersion and non-linear outputs of the SiPMs, an energy correction mechanism is needed. Yet, lack of the information of each crystal's energy will introduce large energy correction error for the ICS events. For this issue, an online energy correction mechanism implemented on a Kintext-7 Field Programmable Gate Array (FPGA) device is presented in this paper. Experiments in the laboratory were performed using an 8 ×8 segmented LYSO crystals coupled with an 8 ×8 SiPM (J-series, from ON Semiconductor) array which is under $^{22}$Na point source excitation. Test results indicate that both the energy of the non-ICS and ICS events can be precisely corrected and the energy resolution is better than 12 %. We also applied this method to an actual clinical PET scanner under a $^{68}$Ge line source to verify its multi-channel reliability.

# Contents

## 1. Introduction

Positron emission computed tomography (PET) based on silicon photomultipliers (SiPMs) has been widely used as an advanced nuclear medicine imaging technique which is with low operation bias voltage ($V_{bias}$), insensitive to magnetic fields and low transit time spread performance [1-3]. Compared with traditional photomultiplier tube (PMT), the compact size of the SiPM allows direct one-to-one coupling between the scintillation crystal and the photo sensor, yielding better timing and energy resolutions [4]. However, the one-to-one coupling scheme requires a huge volume of readout and processing electronics. Although some highly integrated application specific integrated circuits (ASICs) have been presented with the capacity to measure the deposited energy within each crystal inspired by an annihilation photon [5-9], in many low-cost and experimental applications, a front-end multiplexer with position decoder is a common choice to decrease the volume of readout electronics and simplify the system[10-12]. However, as the dimensions of the discrete scintillator elements reduce for better spatial resolution performance, gamma ray inter-crystal scatter (ICS) will increase and the corresponding crystal mispositioning may deteriorate the detector intrinsic resolution in this type system [13]. In addition, the gain dispersion and non-linear outputs of the SiPMs would place some energy events outside of the energy window, resulting in rejected events, so an energy correction mechanism is needed [14]. For the non-ICS events, an annihilation photon is registered in a single crystal and it is simple to correct the energy according to the saturation model of the single SiPM. Yet, for the ICS events, the energy of an annihilation photon are deposited in several crystals, and for the readout electronics based on the multiplexer, lack of the information of each crystal's deposited energy will introduce large energy correction error. [15]

For this issue, combined with the Monte Carlo simulations, this paper presents a simple and precise online energy correction mechanism applied in the one-to-one coupling PET detector based on a Kintex-7 Field Programmable Gate Array (FPGA) device. A multiplexer based front-end prototype was fabricated, and experiments in the laboratory were performed using an 8 × 8 segmented LYSO crystals coupled with an 8 × 8 SiPM (J-series, from ON Semiconductor) array which is under $^{22}$Na point source excitation. Test results indicate that both the energy of the non-ICS and ICS events can be precisely corrected and the energy resolution is better than 12%. To verify its multi-channel reliability, we also applied this method to an actual clinical PET scanner under $^{68}$Ge line source. This work can be employed to furtherly improve the sensitivity performance of our product.

This article is organized as follows. Section-2 introduces the front-end electronics prototype, the position decoder as well as the energy correction method. The evaluation results are introduced in section-3, including the module experiments in the laboratory and tests on an actual PET scanner. A brief summary of this work and the discussion about the improvements at next stage are given in Section-4.

## 2. Design and implementation

### 2.1 Front-end electronics

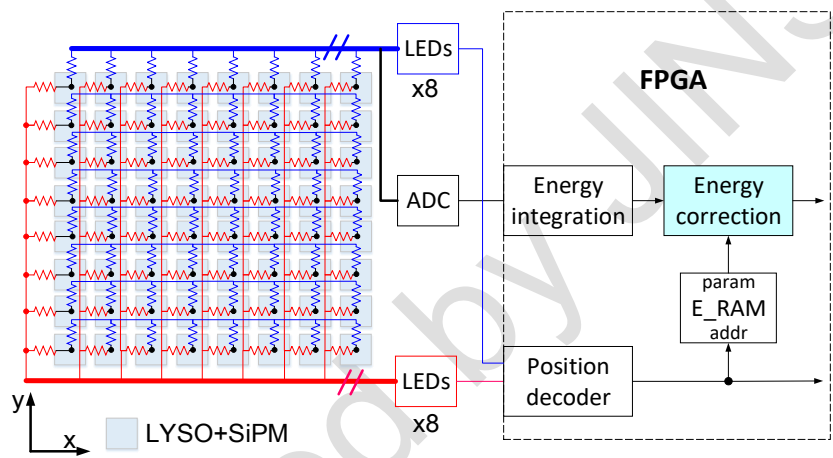

**Figure 1** Architecture of a setup prototype to demonstrate inter-crystal Compton scatter energy correction procedure.

To demonstrate the energy correction procedure, a setup prototype based on a resistance symmetric multiplexer has been conducted [16], as shown in Figure 1. An 8×8 LYSO array (crystal size: 4 mm × 4 mm × 15 mm) is one-to-one pixel coupled to 8×8 SiPMs, which outputs 64-channel current analogue signals from anodes. Each channel is split to two pathways by two resistors (1 k) and a symmetric multiplexing circuit ties the 8-X and 8-Y coordinate signals, separately. So, the number of the readout channels is reduced from 64 to 16. By summing the 8-X signals, the charge for each incident 511-keV photon can be collected, and then digitized by a 50 Msps analog-to-digital converter (ADC, AD9222 from ADI Inc.). The energy information (E) is then calculated by the following energy integration processor implemented in the FPGA. In addition, all the 8-X and 8-Y signals feed 16 leading edge discriminators (LEDs) which provide digital pulses used in the FPGA to provide the index of the hit crystal.

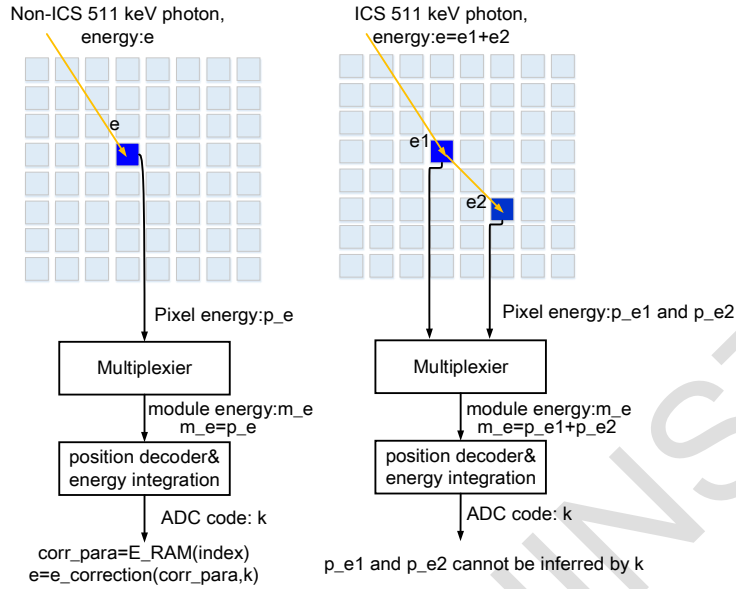

**Figure 2** An example of the energy correction issue for the two-crystal ICS events in the presented setup prototype. For the non-ICS events, the position decoder can infer the index of the hit crystal and the final acquired ADC code (k) can be used to linearly infer the output energy of the inspired single SiPM (p_e). According to the correction parameters of the hit crystal pre-stored in the RAM, the energy of the primary gamma photon can be calculated. For the ICS events, due to the multiplexer, we cannot get the deposited energy of each hit crystal and another different energy correction mechanism should be designed.

As mentioned, the gain dispersion and non-linear outputs of the SiPMs would place some energy events outside of the energy window (425 keV ~ 650 keV in our setup), resulting in rejected events and sensitivity performance deterioration. Therefore, it is necessary to conduct an efficient and precise energy correction. For the non-ICS events, an annihilation photon is registered in a single crystal and once the index of the hit crystal is determined, it is simple to correct the energy by a correction parameters look-up table (LUT) pre-stored in an internal Random Access Memory (RAM) according to the saturation model of the SiPM. However, for the ICS events, two or more crystals are inspired, and we can only get the summing energy of the whole module (m_e), as shown in Figure 2. Thus, we cannot correct the energy for each single crystal, and another energy mechanism should be designed.

The aim of this work is to design and evaluate a simple, precise and fast FPGA based online energy correction algorithm.

## 2.2 Position decoder

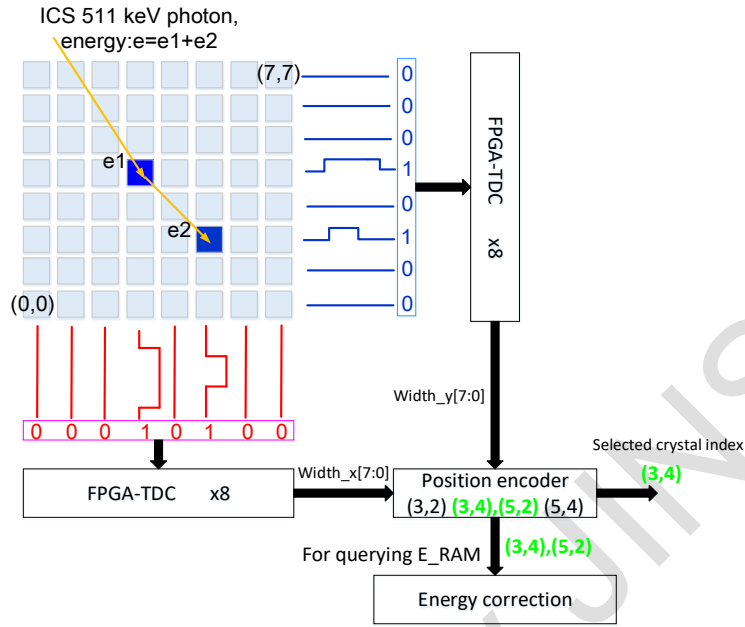

**Figure 3.** The position decoder for two-crystal ICS events.

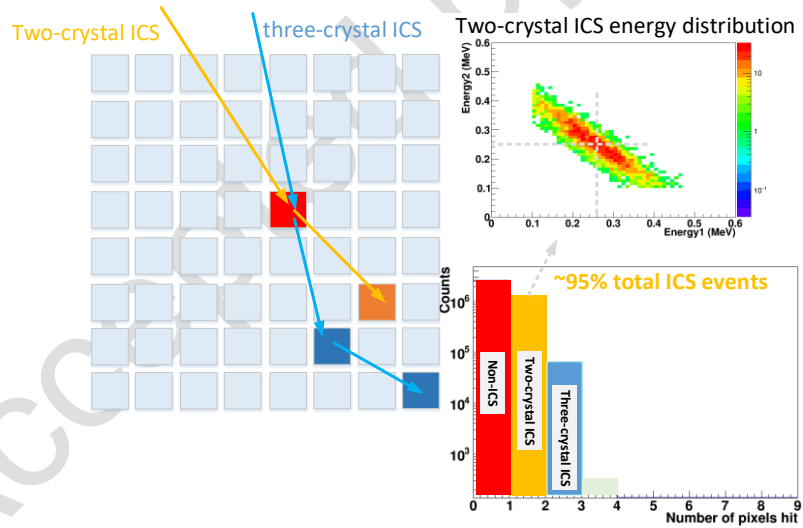

**Figure 4** ICS Monte Carlo simulation results.

Aforesaid, the 8-X and 8-Y position pulses are fed into the position decoder on FPGA to get the index of the hit crystal. For the non-ICS events, only one X hit and Y hit are triggered and the position can be easily decoded. For the ICS events, two or more X (or Y) hits are triggered, and we will get at least two possible indexes. This misposition certainly deteriorates the system spatial resolution. Monte Carlo simulation based on Gate presents that in our system, almost all (~95%) of the ICS events at around 100 keV energy threshold occur between two crystals (two-crystal ICS) and the deposited energy distribution of the two-crystal ICS event is shown in Figure 4. Therefore, in our design, we only decode the position of the two-crystal ICS events. As shown in Figure 3, in the case of two X hits and two Y hits, there will be four decoded index possibilities. To get more precise decoding results, the pulse width of each signal is measured by a time-digital converter implemented in the FPGA (FPGA-TDC) as there is a strong correlation between the pulse width and the energy deposited in the SiPM. According to the measured pulse width, we can combine the X and Y index and choose the crystal deposited with largest energy as the final decoded position. Besides, the two decoded crystal indexes and the ICS events flag are also employed to query the energy correction parameters pre-stored in the E_RAM, whose details are presented in next sub-sections. Specially, for the case that two exactly same hits occur in X and Y axis, there would be four possible decoded position results and we just simply select two randomly.

For an actual clinical PET scanner, we need to implement lots of above TDC in an FPGA device. Therefore, we use the simple multi-phase clock FPGA-TDC to guarantee the source usage tolerance [17]. We layout 192 channel FPGA-TDCs working at 400 MHz sampling clocks with 8 different clock phases(0, 45°, 90°, 135, 180°, 225°, 270°, 315°). So the timing bin size is 312.5 ps. Tests indicate that the measured pulse width at 511 keV is larger than 200 ns, so theoretically the resolution of the FPGA-TDC is good enough.

**2.3 Energy correction algorithm**

Researches has proved that the non-linearity of the SiPM outputs can be corrected according to the below saturation models [18-19].

$$p = -N \cdot [ln(1 - \frac{B \cdot k}{N})]$$

$$E(keV) = p \cdot \varepsilon \qquad (1)$$

where $p$ is the number of photons that would have been detected without saturation; $k$ is the measured energy integration results (ADC code); $B \cdot k$ is the number of triggered microcells with saturation; $N$ is the limited number of microcells in an SiPM; E(keV) and $p \cdot \varepsilon$ are the corrected energy.

Therefore, we have to use at least three radioactive sources to fit Equation-1 and get the correction parameters ($N, B, \varepsilon$) stored in the E_RAM. To simplify it, we transform the saturation model as below.

$$E(keV) = \varepsilon \cdot N \cdot [ln(\varepsilon \cdot N) - ln(\varepsilon \cdot N - \varepsilon \cdot B \cdot k)] = n \cdot [ln(n) - ln(n - b \cdot k)]$$

(2)

where $n = \varepsilon \cdot N$ and $b = \varepsilon \cdot B$.

Therefore, for the non-ICS events, we can use at least two radioactive sources to fit Equation-2 and get the correction parameters ($n, b$) stored in the E_RAM. However, for the two-crystal ICS events, as mentioned before, two crystal indexes and two-set energy correction parameters ($n_0, b_0$) and ($n_1, b_1$) are acquired, so we cannot simply use Equation-2 to correct the energy.

For the two-crystal ICS events, the ADC codes ($k_0$ and $k_1$) can be expressed as below.

$$k_0 = \frac{n_0 \cdot [1-exp(-\frac{E_0}{n_0})]}{b_0} \quad k_1 = \frac{n_1 \cdot [1-exp(-\frac{E_1}{n_1})]}{b_1} \quad (3)$$

According to the datasheet of the J-series 4 mm × 4 mm SiPM and the LYSO crystal, the average limited number of microcells in a SiPM is around 9260, and the light yield of the crystal is around 27000 Ph/MeV. In addition, as Figure 4 shown, it can be approximately considered that the 511 keV photons are equally divided between the two crystals for the two-crystal ICS events [20]. Therefore, for the two-crystal ICS events, the inspired microcells at 256 keV for each SiPM is fewer than the total microcells. That is to say that $\frac{E}{n}$ ($= \frac{p}{N}$) is less than 1. Thus, we can try to simplify Equation-3 by Taylor approximation. For x < 1, the range of the exp(-x) is:

$$1 - x < exp(-x) < \frac{1}{1+x} \quad (4)$$

To decrease the calculation error, we plotted the below four functions and their relative errors with exp(-x), as shown in Figure 5.

$$y_1 = 1 - x$$
$$y_2 = \frac{1}{1 + x}$$
$$y_3 = \frac{1}{2} \cdot [(1 - x) + \frac{1}{1 + x}]$$
$$y_4 = \frac{1}{2}(y_2 + y_3) \quad (5)$$

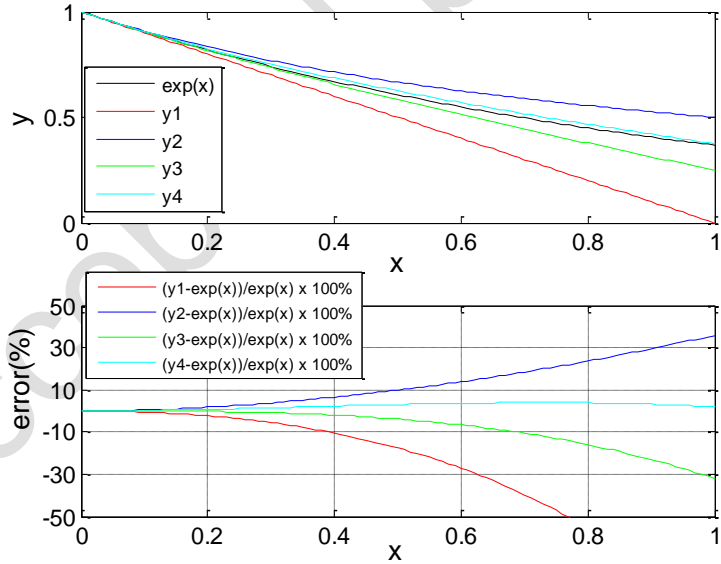

**Figure 5** plotted four approximate functions and their relative errors with exp(-x)

Considering the calculation precision and complexity, we choose $y_3$ as the approximate function to simplify the Equation-3. In addition, for the same SiPM and crystal configuration, $n_0$ is approximately equal to $n_1$, as shown in Equation-6.

$$E_0 \approx E_1 \approx \frac{1}{2}E, \qquad n_0 \approx n_1 \qquad (6)$$

Finally, combined with Equation-3, Equation-5 and Equation-6, the energy can be corrected as the below Equation-7. By the measured summing ADC codes ($k$), we can approximately corrected the ICS energy by the two hit crystals' non-ICS correction parameters ($n_0,b_0$) and ($n_1,b_1$).

$$E = \frac{1}{k \cdot \left(\frac{1}{b_0}+\frac{1}{b_1}\right)-\frac{1}{n_0}} + \frac{1}{k \cdot \left(\frac{1}{b_0}+\frac{1}{b_1}\right)} \qquad (7)$$

Thus, the energy correction algorithm can be expressed as Equation-8. We can use at least two radiative sources to get the non-ICS correction parameters ($n, b$) of each crystal, and then choose the correction equation by the ICS event flag from the position decoder.

$$E = \begin{cases} n \cdot [ln(n) - ln(n - b \cdot k)], & non-ICS\ events \\ \frac{1}{k \cdot \left(\frac{1}{b_0}+\frac{1}{b_1}\right)-\frac{1}{n_0}} + \frac{1}{k \cdot \left(\frac{1}{b_0}+\frac{1}{b_1}\right)}, & ICS\ events \end{cases} \qquad (8)$$

### 2.4 FPGA based online energy correction

Based on the energy correction algorithm (Equation-8), a Verilog-HDL based energy online correction processor has been implemented on Kintex-7 FPGA. The main task is to calculate the Logarithmic ($ln(x)$) and Reciprocal ($1/x$) functions on the premise of low latency, low resource usage and high stability.

We use two internal read only memories (ROMs) as function LUTs to calculate the Logarithmic and Reciprocal functions. This method is with much lower calculation latency than floating float operation and easy to achieve [21]. To guarantee the calculation precision over the large range, the Logarithmic ROM (L_ROM) is designed $4096 \times 14$ bits to record the function $2^{12} \times ln(x)$, and the Reciprocal ROM (R_ROM) is designed to $4096 \times 20$ bits to record the function of $2^{20} \times 1/x$. For the input variable larger than 4096, we use the below bit-shift transformation to normalize it within the range of 1 to 4096. This can help us save lots of memory resources.

$$E = \begin{cases} n \cdot \{2^{12} \cdot ln(n \gg \lambda) - 2^{12} \cdot ln[(n - b \cdot k) \gg \lambda]\} \gg 12, non-ICS\ events \\ \dfrac{1 \gg \lambda_1}{\left[k \cdot \left(\dfrac{1 \gg \lambda_{b0}}{b_0 \gg \lambda_{b0}} + \dfrac{1 \gg \lambda_{b1}}{b_1 \gg \lambda_{b1}}\right) - \dfrac{1 \gg \lambda_{n0}}{n_0 \gg \lambda_{n0}}\right] \gg \lambda_1} + \dfrac{1 \gg \lambda_2}{\left[k \cdot \left(\dfrac{1 \gg \lambda_{b0}}{b_0 \gg \lambda_{b0}} + \dfrac{1 \gg \lambda_{b1}}{b_1 \gg \lambda_{b1}}\right)\right] \gg \lambda_2}, ICS\ events \end{cases}$$
(9)

where $\lambda, \lambda_{b0}, \lambda_{b1}, \lambda_{n0}, \lambda_1, \lambda_2$ are normalization parameters.

The structure of implemented algorithm is shown in Figure 6. The simulated resource usage and calculation latency performance per channel based on the Vivado tool is listed in Table 1.

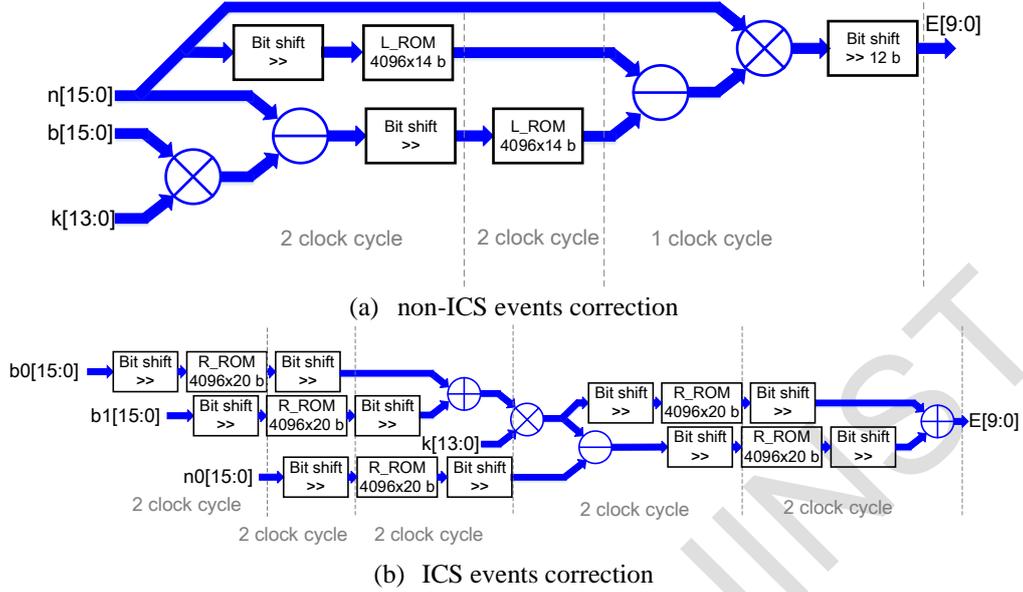

(a) non-ICS events correction

(b) ICS events correction

**Figure 6** Energy correction implemented on FPGA.

**Table 1** the resource usage and calculation latency performance per channel

| item | Simulation results | Utilization on XS7K325TFFG90 Kintex-7 /% |
|---|---|---|
| Block memory usage | 6.5 | 1.46 |
| DSP usage | 3 | 0.36 |
| Clock frequency | 100 MHz | -- |
| Latency | non-ICS events 5 cycles | -- |
| | ICS events 10 cycles | -- |

In summary, the whole structure of the digital signal processor is shown in Figure 7.

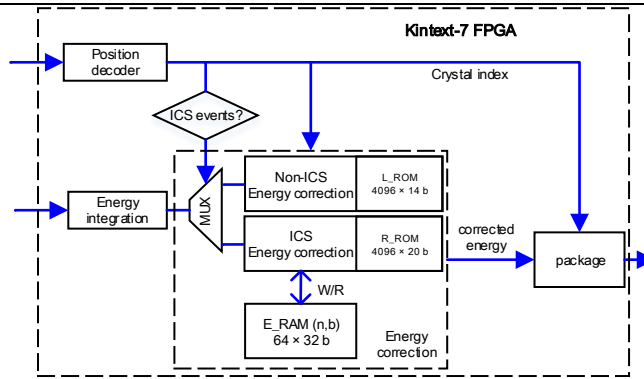

**Figure 7** whole structure of the digital signal processor

## 3. Performance

Test platform as shown in Figure 8 has been set up in the laboratory. An 8 × 8 scintillator array with a pixel size of 4.0 mm (pitch size: 4.2 mm) and thickness of 15 mm is coupled with an 8 × 8 SiPM array with a total area of 33.8 × 33.5 mm$^2$ (pitch size: 4.2 mm) which is under $^{22}$Na point source excitation (~50 μCi, around 15 cm distance). The outputs of the SiPMs are processed by the multiplexing board and then fed into the front-end board via flat cable to achieve the position and energy calculation. The front-end board consisting of one Kintex-7 FPGA device can receive the signals from 12 SiPM arrays. The front-end board sends the measurement data to data-acqusion (DAQ) via fibers, and the DAQ communicates with the computer through a PCIE interface. The operating bias voltage for all SiPMs on one module is 29.5 V, and all the LYSO crystals have been polished and later covered by reflective material (Teflon). The LYSO and the SiPM are pasted with optical grease (BC630).

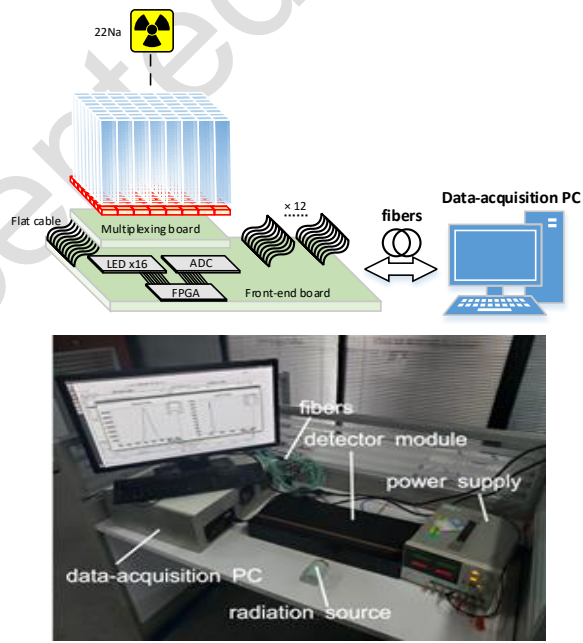

**Figure 8** platform in test

## 3.1 module test

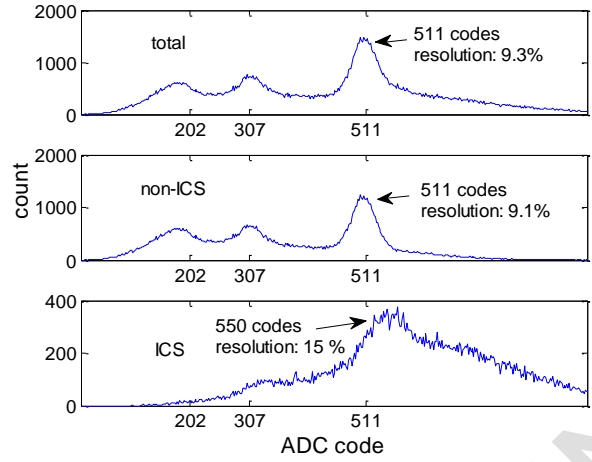

**Figure 9** the energy spectrum of total, non-ICS and ICS events before the energy nonlinearity correction

Before energy correction, the energy spectrum of all the crystals in the module is presented in Figure 9. To facilitate observation, all the energy peaks are linearly normalized to 511 codes. The results show that the ICS events are with a larger gamma energy peak (550 codes) and bad resolution due to the error on the position and the saturation of the SiPMs. This would cause data loss within the energy window. The energy resolution for the non-ICS events is 9.1%.

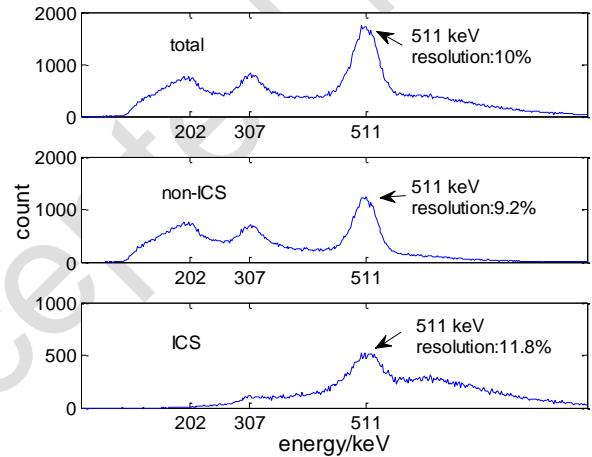

**Figure 10** the energy spectrum of total, non-ICS and ICS events after the energy nonlinearity correction

We use the natural radiation of the LYSO (202 keV, 307 keV) and the 511 keV of $^{22}$Na to calibrate the energy correction parameters. The energy spectrum of the module after correction is shown in Figure 10. It shows that the gamma energy peak of the ICS events is precisely corrected at 511 keV and with a better resolution (~11.8 %). The experiment results indicate that the presented energy correction processor shows good performance. The energy resolution for the non-ICS events is 9.2%.

## 3.2 clinical PET scanner test

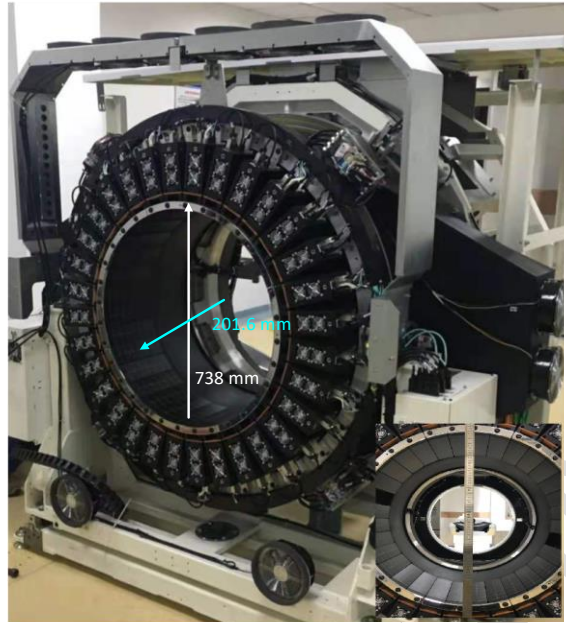

**Figure 11** the Minfound scintcare Lynx series PET scanner.

In order to verify the energy correction processor's multi-channel reliability, we tested it on a clinical PET scanner (Scintcare Lynx series, from Minfound Inc.) under a $^{68}$Ge line source (around 500 μCi) placed at centre as shown in Figure 11. This PET scanner consists of 32 detector sections and 384 LYSO crystals one-to-one pixel coupled with the J-series 4 mm × 4 mm SiPMs. The limited field of vision (FOV) is 738 mm. Figure 12 shows the energy spectrum of the whole scanner and the total energy resolution is 10.1%. The energy spectrum of the ICS events is precisely corrected.

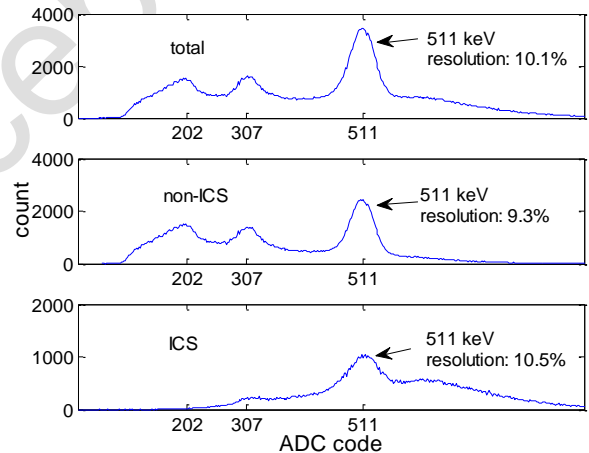

**Figure 12** the energy spectrum of total, non-ICS and ICS events after the energy nonlinearity correction (whole PET scanner)

## 4. Summary

This paper presents a simple online FPGA based energy correction processor for the multiplexed readout of the one-to-one SiPM-LYSO coupling PET detector. According to the saturation model of SiPM outputs, an approximate energy correction formula for the ICS events is derived and implemented on the FPGA. Experiments in the laboratory were performed using an 8 × 8 segmented LYSO crystals coupled with an 8 × 8 SiPM (J-series, from ON Semiconductor) array which is under $^{22}$Na point source excitation. Test results indicate that both the energy of the non-ICS and ICS events can be precisely corrected and the energy resolution is better than 12 %. We also applied this method to an actual clinical PET scanner under a $^{68}$Ge line source to verify its multi-channel reliability. NEMA tests indicate that this energy correction method can increase the sensitivity performance by about 10%. The next step of the research would be finding a similar method for the light-share detector and the timing correction for the ICS events.

## 5. Acknowledgement

All authors declare that they have no known conflicts of interest in terms of competing financial interests or personal relationships that could have an influence or are relevant to the work reported in this paper.

The authors would like to thank the State Key Laboratory of Particle Detection and Electronics for the radiation experiment platform.

## References

[1] K. Wagatsuma et al., "Comparison between new-generation SiPM-based and conventional PMT-based TOF-PET/CT," *Physica Medica*, vol. 42, pp. 203–210, Oct. 2017, doi: 10.1016/j.ejmp.2017.09.124

[2] A. D. Guerra et al., "Silicon photomultipliers (SiPM) as novel photodetectors for PET," *Nucl. Instrum. Methods Phys. Res. A*, Accelerators Spectrometers Detectors Assoc. Equip., vol. 648, pp. S232–S235, Aug. 2011, doi: 10.1016/j.nima.2010.11.128

[3] F. Powolny et al., "Time-based readout of a silicon photomultiplier 490 (SiPM) for time of flight positron emission tomography (TOF-PET)," *IEEE Trans. Nucl. Sci*., vol. 58, no. 3, pp. 597–604, Jun. 2011, doi: 10.1109/TNS.2011.2119493.

[4] Surti S , Karp J S . Impact of event positioning algorithm on performance of a whole-body PET scanner using one-to-one coupled detectors[J]. *Physics in Medicine & Biology*, 2018.

[5] Francesco A D , Bugalho R , Oliveira L , et al. TOFPET2: a high-performance ASIC for time and amplitude measurements of SiPM signals in time-of-flight applications[J]. *Journal of Instrumentation*, 2016, 11(03):C03042.

[6] Chen, Huangshan, Harion, et al. STIC3-Silicon Photomultiplier Timing Chip with picosecond resolution[J]. *Nuclear Instruments and Methods in Physics Research, Section A. Accelerators*, Spectrometers, Detectors and Associated Equipment, 2015, 787:284-287.

[7] Yuan Z , Briggl K , Chen H , et al. KLauS: A Low-power SiPM Readout ASIC for Highly Granular Calorimeters[C]// 2019 IEEE Nuclear Science Symposium and Medical Imaging Conference (NSS/MIC). IEEE, 2019.

[8] D Schug, Nadig V , Weissler B , et al. Initial Measurements with the PETsys TOFPET2 ASIC Evaluation Kit and a Characterization of the ASIC TDC[J]. *IEEE Transactions on Radiation & Plasma Medical Sciences*, 2018:1-1.

[9] Sanchez D, Gomez S, Mauricio J, et al. HRFlexToT: A High Dynamic Range ASIC for Time-of-Flight Positron Emission Tomography[J]. *IEEE Transactions on Radiation and Plasma Medical Sciences*, 2021, PP(99):1-1.

[10] E. Downie, X. Yang, and H. Peng, "Investigation of analog charge multiplexing schemes for SiPM based PET block detectors," *Phys.Med. Biol.*, vol. 58, no. 11, pp. 3943–3964, 2013, doi: 10.1088/0031 9155/58/11/3943.

[11] S. Lee, Y. Choi, J. Kang, and J. H. Jung, "Development of a multiplexed readout with high position resolution for positron emission tomography," *Nucl. Instrum. Methods Phys. Res. A, Accelerators Spectrometers Detectors Assoc. Equip.*, vol. 850, pp. 42–47, Apr. 2017, doi: 10.1016/j.nima.2017.01.026.

[12] H Kim, Kao C M, Hua Y, et al. Multiplexing Readout for Time-of-Flight (TOF) PET Detectors Using Striplines[J]. *IEEE Transactions on Radiation and Plasma Medical Sciences*, 2021, PP(99):1-1.

[13] Shao Y, Cherry S R. A study of inter-crystal scatter in small scintillator arrays designed for high resolution PET imaging[J]. *IEEE Transactions on Nuclear Science*, 1996, 43(3):1938-1944.

[14] Grazioso R, D Henseler, Zhang N. Energy correction for one-to-one coupled radiation detectors having non-linear *sensors*[J]. 2014.

[15] E Yoshi da, Obata F, Kama Da K, et al. Development of Single-Ended Readout DOI Detector With Quadrisected Crystals[J]. *IEEE Transactions on Radiation and Plasma Medical Sciences*, 2020, PP(99):1-1.

[16] Kwon, Sun, Il, et al. Signal encoding method for a time-of-flight PET detector using a silicon photomultiplier array[J]. *Nuclear Instruments & Methods in Physics Research*, 2014.

[17] Dong X, Ma C, Zhao X, et al. A high resolution multi-phase clock Time-Digital Convertor implemented on Kintex-7 FPGA [J]. *Journal of Instrumentation*, 15 T11005.

[18] Ma C, Dong X, Yu L, et al. Design and evaluation of an FPGA-ADC prototype for the PET detector based on LYSO Crystals and SiPM arrays[J]. *IEEE Transactions on Radiation and Plasma Medical Sciences*, 2021, PP(99):1-1.

[19] C. Degenhardt et al., "The digital silicon photomultiplier—A novel sensor for the detection of scintillation light," in Proc. IEEE Nucl. Sci. Symp. Conf. Rec. (NSS/MIC), Orlando, FL, USA, 2009, pp. 2383–2386, doi: 10.1109/NSSMIC.2009.5402190

[20] Geng F, Ivan A, Hua Q. Recovery of inter-crystal compton scattering events for sensitivity improvement of sub-250 ps TOF-PET detector[C]// IEEE Nuclear Science Symposium. IEEE, 2017.

[21] Fu H, Mencer O, Luk W. FPGA Designs with Optimized Logarithmic Arithmetic[J]. *IEEE Transactions on Computers*, 2010, 59(7):1000-1006.